\documentclass{aastex}
\usepackage{spr-astr-addons}
\usepackage{url}
\RequirePackage{color}

\begin{document}
\title{Phase speed  and  frequency-dependent damping of longitudinal intensity oscillations in coronal loop  structures observed with  AIA/SDO}
\slugcomment{Not to appear in Nonlearned J., 45.}
\shorttitle{Short article title}
\shortauthors{Autors et al.}
\author{A.Abedini\altaffilmark{1}}
\affil{Department of Physics, University of Qom,  Qom University Blvd Alghadir, P.O. Box  3716146611, Qom, I. R. Iran.\\
email:a.abedini@qom.ac.ir}
\begin{abstract}
Longitudinal intensity oscillations along coronal loops that are interpreted as signatures of  magneto-acoustic waves are observed frequently  in different coronal structures.
The aim of this paper is to estimate the physical parameters of the slow waves and the quantitative dependence of these parameters on their frequencies in the solar corona loops that are situated above active regions with the  Atmospheric Imaging Assembly (AIA) onboard Solar Dynamic Observatory (SDO).
 The observed data on 2012-Feb-12, consisting  of  300  images with an interval of 24 seconds in the  171 $\rm{\AA}$ and 193 $\rm{\AA}$ passbands is analyzed for evidence of
 propagating features as slow waves along the loop structures.
  Signatures of  longitudinal intensity oscillations that are damped rapidly as they travel along the loop structures were found, with periods  in the range of  a few minutes to few tens of minutes.
  Also, the projected (apparent) phase  speeds, projected damping  lengths, damping times and
 damping qualities  of   filtered  intensities  centred on the dominant frequencies  are measured in the range of  $\rm{C_s}\simeq 38-79~ \rm {km~s^{-1}}$,  $\rm{L_d}\simeq 23-68 ~\rm{Mm }$,  $\rm{\tau_d}\simeq 7- 21 ~\rm {min}$ and $\rm{\tau_d/P}\simeq 0.34- 0.77$, respectively.
The theoretical  and observational results of this study indicate that the damping  times and damping  lengths  increase  with increasing the oscillation  periods, and   are   highly sensitive  function of oscillation  period, but the projected  speeds and the damping qualities are not very sensitive to the  oscillation periods. Furthermore, the magnitude values of  physical parameters  are   in good agreement  with the prediction of the theoretical dispersion relations of high-frequency  MHD waves ($>1.1~ \rm{mHz}$) in a coronal plasma with electron number density in the range of
$\rm{n_e}\simeq 10^{7} - 10^{12} ~\rm{cm^{-3}}$.
\end{abstract}
\section{Introduction}
The observational results  of   satellites  such as
Yohkoh, Solar and Heliospheric Observatory  (SoHO),  Transition Region And Coronal  Explorer (TRACE), Hinode, Solar Terrestrial Relations Observatory (STEREO) and SDO indicate that there are now clear examples of low-amplitude quasi-periodic intensity oscillations  with long-oscillation  periods in different coronal structures (see, {\it e.g.}, Ofman et al. \citeyear{Ofman1997},~\citeyear{Ofman1999}; DeForest and Gurman  \citeyear{Deforest1998};
 Berghmans et al. \citeyear{Berghmans1999}; De Moortel et al. \citeyear{DeMoortel2000},~\citeyear{DeMoortel2009}; Mariska \citeyear{Mariska2006}; McEwan and De Moortel \citeyear{McEwan2006}; Wang et al. \citeyear{Wang2009}; Mariska and Muglach \citeyear{Mariska2010}).
These low-amplitude  intensity oscillations  may be caused by propagating or standing  slow magneto-acoustic waves.
Propagating intensity oscillations in coronal holes high above the limb were first observed
 by Ofman et al. (\citeyear{Ofman1997}). Deforest and Gurman (\citeyear{Deforest1998}) observed similar propagating intensity oscillations (compressive waves trains)  in the plumes.
 Ofman et al. (\citeyear{Ofman1999}) suggested that these propagating oscillations may be caused  by  magneto-acoustic wave due to their propagating speeds  that
are close to the sound speed in the corona.
Also, intensity oscillations with large Doppler-shift velocities and strong oscillatory damping
detected in hot  coronal loops. These oscillations were interpreted as signatures of longitudinal slow magneto-acoustic mode  excited impulsively in the corona  loops. They had periods from about 7 to 31  minutes and decay times in the range of 6-37 minutes
(see, {\it e.g.}, Ofman and Wang  \citeyear{OfmanW2002};  Kliem et al. \citeyear{Klieme2002}; Sakurai et al. \citeyear{Sakurai2002};  Wang et al. \citeyear{Wang2002a}, \citeyear{Wang2002c}, \citeyear{Wang2003a}, \citeyear{ Wang2003b}, \citeyear{Wang2011};
Banerjee et al. \citeyear{Banerjee2007}; Erd{\'e}lyi et al. \citeyear{Erdelyi2008}).
There have been many theoretical studies examining  the standing and propagating  longitudinal  slow magneto-acoustic waves  in coronal loop structures.
 Theoretical studies investigating the damping of the slow waves have concentrated on  the effects of thermal conduction, compressive viscosity, optically thin radiation, gravitational stratification and magnetic field divergence.
In general, thermal conduction is found to be the dominant mechanism for dissipation of slow waves in the corona (see, {\it e.g.},
Porter et al. \citeyear{porter1994a}; Roberts \citeyear{Roberts2000}, \citeyear{Roberts2006};  Nakariakov et al. \citeyear{Nakariakov2000}, \citeyear{Nakariakov2005};
Tsiklauri and Nakariakov, \citeyear{Tsiklauri2001};   Ofman et al. \citeyear{Ofman2000};   De Moortel and Hood \citeyear{DeMoortelh2003}, \citeyear{DeMoortelh2004a}; Mendoza-Brice{\~n}o et al. \citeyear{Mendoza2004}; Klimchuk et al. \citeyear{Klimchuk2004}; Verwichte et al. \citeyear{VerwichteN2005}, \citeyear{Verwichte2008}; Van Doorsselaere et al. \citeyear{VanDoorsselaere2011};  Abedini and Safari,  \citeyear{Abedini2011};  Abedini et al. \citeyear{Abedini2012}).
 The idea of coronal seismology was first suggested by  Uchida  (\citeyear{Uchida1968},~\citeyear{Uchida1970}).
Then, this idea has been both widely and successfully employed to determine coronal properties and  MHD waves.
For example,  Marsh and Walsh  (\citeyear{Marshw2009}) presented the  three-dimensional observations of coronal slow magneto-acoustic wave propagation. The magnitude of 132 $ \pm $ 9 and 132 $ \pm 11~\rm km~s^{-1}$    was found for coronal longitudinal slow mode speed  with
STEREO  A and STEREO B  observation.
Marsh et al. (\citeyear{Marsh2011}) determined the density profile of the loop system using Hinode observations and measured the  three-dimensional amplitude decay length of the slow wave.
The magnitude of the three-dimensional  decay length of slow wave was found  20 and 27 Mm  for STEREO A and STEREO B  observation, respectively.
Van Doorsselaere et al. (\citeyear{VanDoorsselaere2011}) used observations of a slow MHD wave in the corona to determine for the first time the value of the effective adiabatic index by  the Extreme-ultraviolet Imaging Spectrometer (EIS) on board Hinode.
The  magnitude of the effective adiabatic index  for slow wave was  measured $1.01\pm 0.02$ and  found  that the thermal conduction is  dominant mechanism for dissipation of slow waves in the corona.
Yuan and Nakariakov (\citeyear{Yuan2012}) analyzed  the quasi-periodic Extreme Ultra-
Violet (EUV) disturbances propagating at a coronal fan-structure of active region by AIA/SDO in 171 $\rm{{\AA}}$.
They designed cross-fitting technique, 2D coupled fitting  and best similarity match to measure the apparent phase speed of propagating disturbances in the running differences of time-distance plots (R) and background-removed and normalised time-distance plots (D).
In this  analysis, the average  apparent phase speed  was measured at  47.6  $\rm{km~s^{-1}} $ and 49.0 $\rm{km~s^{-1}} $  for R and D, with the  corresponding oscillation periods at $ 179.7 \pm 0.2$ s and $179.7 \pm 0.3 $,  respectively.  Threlfall et al.(\citeyear{Threlfall2013})  studied both transverse and longitudinal motion by comparing and contrasting time-distance images of parallel and perpendicular cuts along/across active region fan loops.  The  apparent phase speed  of transverse and longitudinal  oscillations are  found $600 -750~ \rm{km~s^{-1}}(\rm{P}=3- 8~ \rm{min})$ and
  $100 -200 ~\rm{km~s^{-1}}(\rm{P}=6-11 ~\rm{min})$  along loop structures and above the limb, respectively.
  Moreover, the propagation speeds of   waves in the  sunspots, non-sunspots, loop  structures and polar regions were studied  by many authors (see, {\it e.g.}, Kiddie et al. \citeyear{Kiddie2012};  Uritsky et al.  \citeyear{Uritsky2013}).
Recently, Krishna Prasad et al. (\citeyear{Krishna2014})  studied both theoretically and observationally the quantitative dependence of projected damping length of the magneto-acoustic wave on its frequency.
They found that damping length on loop structures over a sunspot and an on-disk plume
like structure increases with  oscillation period. The plume and interplume structures
at the south pole damping length decrease with oscillation period.
Although damping  of  slow waves has been studied extensively in
different coronal structures, but, studies on the periodicity dependence of their damping are relatively limited.
In this paper, a new  method is applied to investigate  the nature of
longitudinal intensity fluctuation in active-region coronal loop structures   observed with AIA/SDO.
The averages and ranges of both the apparent and real physical parameters of longitudinal intensity fluctuation for six different paths along the corona loops such as   projected   damping length,  projected phase speed, oscillation periods, damping time and damping quality  has been extracted  in 171$\rm {\AA}$ and 193$\rm {\AA}$  passbands.
Moreover,  both theoretically and observationally the quantitative dependence of damping  of the intensity fluctuation  on its frequency is studied.
The paper is organized as follows.
Section 2 describes the observations.
Section 3 describes method of  data analysis.
Section 4 describes the method of the physical parameters calculation.
Section 5 presents the theoretical considerations.
Finally, discussion and conclusions are given in section 6.
\section{ Observations }
\begin{figure}    
    \centerline{\includegraphics[width=.5\textwidth,clip=]{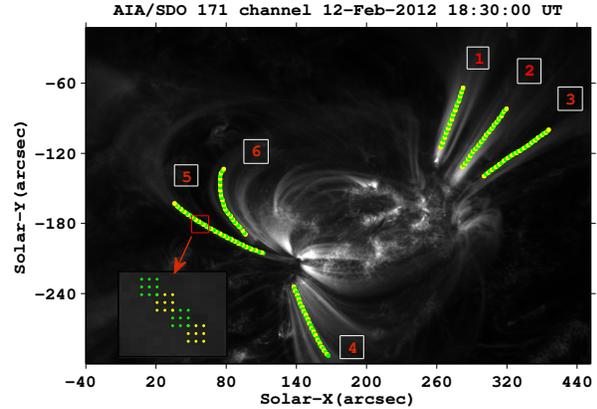}
              }
              \caption{ A snapshot of  the active region (AR11416), observed by AIA/SDO in 171$\rm {\AA}$ on 2012-Feb-12, 18:30 UT. Locations of the selected coronal loop paths  that are   situated above this active region, are also marked.
               The coronal loop structures display intensity  propagation signal.
               In order to obtain the intensity as function of distance along loop segments, these  paths also  are  subdivided  by successive macropixels with  3$\times$3  pixels wide along  the paths. A zoomed in view of a partial segment on loop 1   is shown in the bottom left corner of the Figure.
                     }
                     \vspace{+0.001\textwidth}
   \label{AR11416}
   \end{figure}

The observations of interest for this study are based upon AIA/SDO  data.
The AIA data consist of high cadence (12 s)  images of the solar corona in 10 UV and EUV wavelengths.
The images must be cleaned and calibrated using a number of reduction procedures before the data can be analyzed.
The data used here is at level of 1.5. On the other hand, flat-fielding, co-aligned, filter, vignette, and bad pixel/cosmic-ray corrections  have already been applied on the data. Furthermore, the images have been rescaled to a standard $0.6^{"}$ plate scale, and  have   been rotated so that solar north is up in the image.
In order to obtain the  physical properties  of longitudinal intensity fluctuation in the large active -region loops,
we need  an  active region  loop system which supports slow-mode wave propagation.
The data under analysis here is taken on 12 February 2012, from 18:30 UT  until 20:30 UT
and it consists of  a series of $800\times460$ pixel, Sun-centered, subfield images on the 171$\rm {\AA}$ (Fe IX) and 193$\rm {\AA}$	 (Fe XIV)  passbands  with a  time-distance 24 s.
All of  coronal loop  structures observed are situated above  an interesting active region; this region is numbered  AR11416 (12 February 2012).
Figure \ref{AR11416} shows an image of AR 11416  at 18:30 UT on which we are concentrating in this paper.
The six different loop segments  in Fig. \ref{AR11416} show the paths we will be looking at in detail  for analysis.
In order to obtain the intensity as function of distance along loop segments, these  paths also  are  subdivided  by macropixels with  3$\times$3  pixels wide along  the loop segments.
A zoomed in view of a partial segment on loop 1   is shown in the bottom left corner of the Fig. \ref{AR11416}.
   \section{Data Analysis}\label{Dataa}
\begin{figure}    
  \centerline{\includegraphics[width=.5\textwidth,clip=]{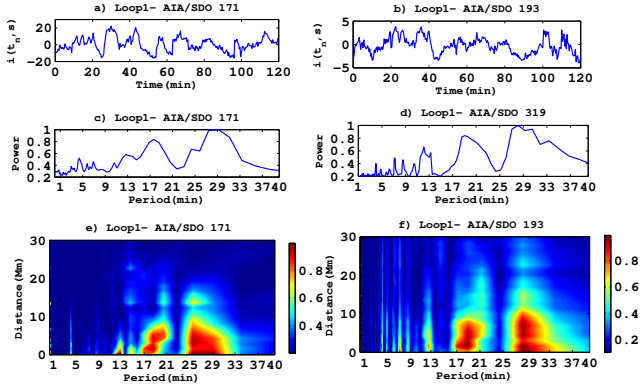}
              } \caption{The variation of background-subtracted intensities with time at the 5th macropixel  along the loop 1 are shown in the top row panels (a) and b)) and,  their   FFT power spectral densities  in the middle row(c) and d)). The Period-distance maps are shown in the bottom row panels for  171$\rm {\AA}$ (left panels), and 193$\rm {\AA}$ (right panels) passbands of AIA/SDO.}
             \vspace{-0.02\textwidth}
   \label{perioddm}
   \end{figure}

In order to investigate the oscillatory nature of the intensity propagations  along the coronal loop system, the marked regions along the path of the loops  are  joined by successive macropixels, 3$\times$3 pixels wide along the loop strands.
The intensity of each macropixel  along the loop  system is integrated and divided into  the number  of pixels at each  macropixel.
A quasi-static background  must be removed from original intensity for enhancing the contrast of intensity variation.
  Following  Yuan and Nakariakov (\citeyear{Yuan2012}), a background intensity of the form
\begin{eqnarray}
&&  I_b(s,t_n)=\sum_{h=-N/2}^{+N/2-1}I(s,t_{n+h})/N,
\label{intensity}
  \end{eqnarray}
is subtracted from time series of intensities, where N is the number of time frames,  $t_n$  represent the time frame index   of the image series ($n=1,~ t_1=0 $ corresponds to the first image, $n=2,~t_2=24 s $ corresponds to the second  image, ...) and   $ s $ determines  the location  of macropixels  along the indicated paths in Fig. \ref{AR11416} (starting at first  sector).
An appropriate background constructed from the 100 time frames (48 min) running average in time is subtracted from the original intensities, because a sufficient enhanced time-distance maps  is found by setting the $N=100$.
It is worth noting  that the periods greater than 48 minutes will be suppressed from power spectra of intensities time series by choosing $N=100$.
\begin{figure}    
   \centerline{\includegraphics[width=.5\textwidth,clip=]{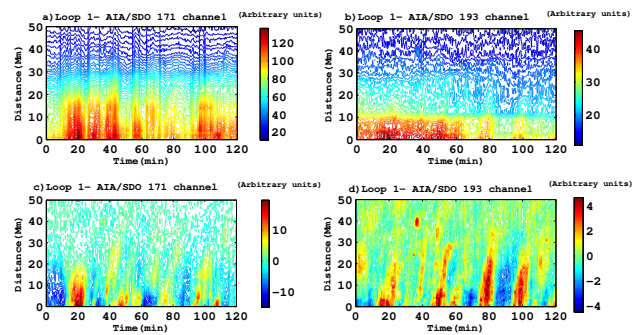}
              }
              \caption{Time-distance maps of loop 1 for original intensity (top row), and for  background-subtracted intensity (bottom row) are shown in 171 (a) and c) panels) and 193 (b) and d) panels) channels of AIA.
                }
                              \vspace{+0.01\textwidth}
   \label{Iandi}
   \end{figure}
For example, time-distance maps of intensity along the loop 1 for 171$\rm{{\AA}}$ and 193 $\rm{{\AA}}$ before (top panels a) and b)) and after (bottom  panels c) and d)) removal of the background intensity are shown in  Fig. \ref{Iandi}.
Intensity as a function of time   at each macropixels of six different
loop segments  which are shown  in Fig. \ref{AR11416} is calculated.
Also, variation of background-subtracted intensities with time at the 5th macropixel  along the loop 1 are shown in the top row panels (a) and b)) of Fig. \ref{perioddm} and,  their  FFT power spectral densities  in the middle row(c) and d)).  Also, the period-distance maps are shown in the bottom row panels (e) and f))  for 171$\rm {\AA}$ (left panels), and 193$\rm {\AA}$ (right panels).
The Fourier  power spectra and period-distance maps of background-subtracted intensity reveal periods in the range of $\rm{P}$ = 8-40 minutes (Fig. \ref{perioddm},  middle and bottom row panels).
The period-distance maps of intensities show that  significance  levels ($>0.5$) are  in the $\rm{P}=12-35\rm{(min)} $  oscillation periods range.
Furthermore, the spectral power densities factor of  intensities  are dominated at some periods between 12 min and 35 min.
 \section{The method of the physical parameters calculation }
Here, the physical parameters of propagating disturbances that are important tools for MHD coronal seismology  are extracted  with a new method.
   So, the Fourier components of the intensity time series  at each macropixel along the loop structure are multiplied with a Gaussian function of the form
  \begin{eqnarray}\label{temp }
a^{'}_j =\sum_{i=1}^{N_f}a_i \exp [-\frac{(\nu_i-\nu_j)^{2}}{2\sigma ^{2}} ],
\end{eqnarray}
where $N_f$  is the number of the Fourier components, $a_i $  one of the Fourier components, $a^{'}_j $ one of the interesting dominant component, $\sigma $ standard deviation and $a^{'}_i $ transformed coefficients, respectively (subscript index i and j represent the rank of the harmonics).
Also, $\nu_i$ and $\nu_j$ represent the frequency  of the ith and jth harmonics.
In the Fourier transformation, the component proportional to the  dominant oscillation period   is selected by using a Gaussian filter with standard deviation $ \sigma\simeq(\nu_{i+1}-\nu_{i-1})/2$ to enhance time-distance maps.
For example, Fig. \ref{phasespeed} shows the time-distance maps of the loop 1  in 171 \rm{{\AA}} (left panels) and 193\rm{{\AA}} (right panels) with the Gaussian filter centered on the dominant oscillation periods of intensity.
The time-distance maps of filtered intensities  clearly show evidence that acoustic waves propagate upwardly into the coronal loop 1.
In the next subsections, the physical parameters of propagating disturbances (slow waves) such as phase speed, damping length, damping time and  damping quality are extracted   by  analyzing the  time series of filtered intensities and their enhanced time-distance maps.
Furthermore, the results are compared with the previous studies.
\subsection{The measurement of the projected phase speed}
\begin{figure}    
   \centerline{\includegraphics[width=.5\textwidth,clip=]{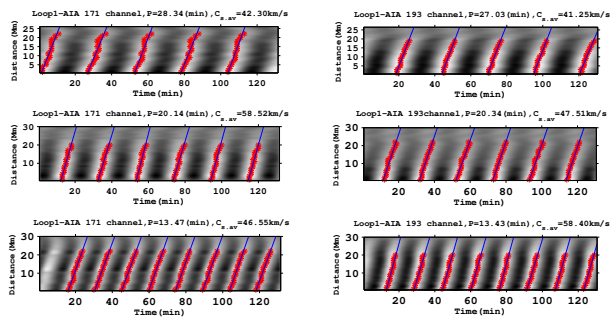}
              }
                        \caption{Quasi-periodic longitudinal features along the loop1 in 171 (left panels) and 193 (right panels) channels of AIA.
                        The location of the points that have a maximum upward displacement  are indicated  with red color $\ast$  signs. The solid blue lines represent a linear function that are fitted to the red star signs, and the gradient  of the solid lines in each panel are used to estimate  average projected speed along loop 1  at a particular period.
                         }
             \vspace{-0.02\textwidth}
   \label{phasespeed}
   \end{figure}

During the last years several methods are used to determine the  speed of a propagating disturbances.
Many authors calculated the speed  of the propagating disturbances  from  the gradient of
the lines by fitting a linear function to the points with similar amplitude fluctuation of unfiltered time-distance maps.
 Also, some authors measured  the speed  of the propagating disturbances by fitting a propagating harmonic wave function to the unfiltered  propagating disturbances (see, {\it e.g.}, De Moortel et al. \citeyear{DeMoortel2000}; Kiddie et al. \citeyear{Kiddie2012}; Yuan et al. \citeyear{Yuan2012}; Threlfall et al. \citeyear{Threlfall2013}).
Here, the time-distance maps of filtered intensities that have periodic  features, are used  to estimate projected phase speeds of propagating disturbances.
For example, Fig. \ref{phasespeed} shows the time-distance of the loop 1  in 171 \rm{{\AA}} (left panels) and 193\rm{{\AA}} (right panels) with the Gaussian filter centered on the dominant oscillation periods of intensity. The points that have a maximum upward displacement  are indicated  with red color star signs. The  solid blue lines  represent a linear function that are fitted to the red star  signs, and the gradient  of  solid lines are used to estimate average projected speeds of propagating disturbances for some particular oscillation periods,
 also the derived speed   for a particular period  have been written in the respective panels of Fig. \ref{phasespeed}.
 It can be seen that oscillation at a particular period propagates with an almost constant speed  along the loop 1  in both passbands.
The averages and ranges of observed  speeds along the six different  paths are compiled in Table \ref{observedresult}.
The averages and ranges of observed  speeds comfortably overlap with previous finding by other authors who analyzed similar observations of propagating fluctuations (see, {\it e.g.}, De Moortel et al. \citeyear{DeMoortel2000}; Marsh et al. \citeyear{Marsh2003}, \citeyear{Marshw2009}; Kiddie et al. \citeyear{Kiddie2012}; Yuan et al. \citeyear{Yuan2012}; Threlfall et al. \citeyear{Threlfall2013};
Sych and  Nakariakov, \citeyear{Sych2014}).
This method has several advantages.
(1) The time-distance maps of filtered intensities clearly show evidence
that acoustic waves propagate upwardly into coronal loops (see Fig.\ref{phasespeed}).
(2) To distinguish the  points  with maximum  displacement (or points with  similar displacement) from these maps  is very much easier  than the time-distance maps of unfiltered intensities,
then this method  can  produce reliable results.
(3) Furthermore, this method reveals   how the  magnitude of physical parameters vary depending on the oscillation periods.

\subsection{The measurement of damping of slow waves  and  the dependence of damping value  on its frequency}
\begin{figure}    
  \centerline{\includegraphics[width=.5\textwidth,clip=]{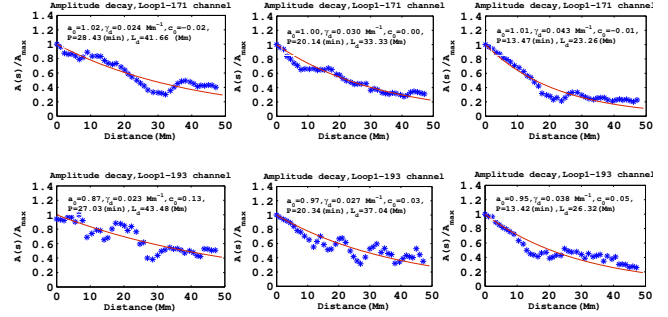}}\caption{Typical examples of the normalized  amplitude decay  profiles $ A(s)/A_{max} $ for loop 1  at $\rm{P}=28.43,~27.03,~20.14,~20.34,~13.14 $ and  $13.42~\rm{min}$  are shown as function of the projected distance. The data  is fitted   by   an exponential function, $A(s)/A_{max} = a_0 \exp(-\gamma_d s)+c_0$ (red lines), and  fit parameters are written  in the top of  the  panels.              Top and bottom panels correspond to the data from 171 and 193 channels respectively.}
                      \vspace{+0.01\textwidth}
   \label{amplituded}
   \end{figure}

\begin{figure}    
  \centerline{\includegraphics[width=.55\textwidth,clip=]{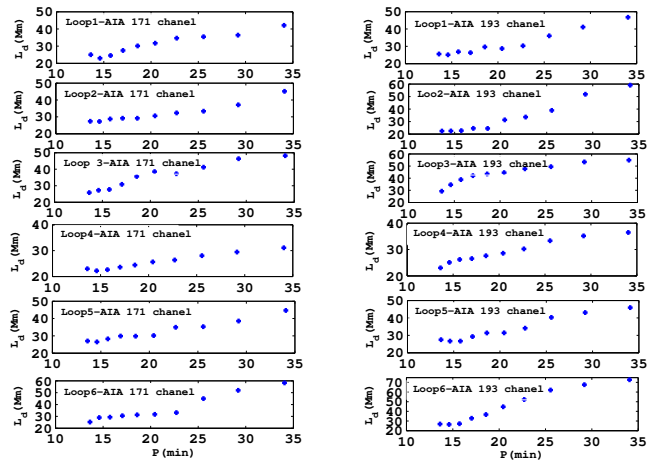}
              }
              \caption{Variation of observed damping lengths $\rm{L_d} $ with period
               in 171 (left panels) and 193 (right panels) channels of \rm{AIA} for 6 loop segments
marked  in the  panel of Fig. 1.}
             \vspace{-0.02\textwidth}
   \label{dampinglengths}
   \end{figure}

Damping of MHD waves have been studied extensively in coronal structures  both theoretically and observationally since their first detection. However, studies on the dependence of damping value  on its frequency are limited.
Fourier power spectral densities  of  background-subtracted intensities show that Fourier components of the intensity time series at each macropixel along the six different  paths are only significant at some specified  period in the range of $\rm{P}= 10 -35$ minutes.
For example, the period-distance maps  constructed for loop 1 in the 171 (bottom left panel) and 193 \rm{{\AA}}
(bottom right panel), are shown in the  Fig. \ref{perioddm}.
To estimate the damping length of MHD waves and the  dependence of damping length on its frequency, the amplitude  at a particular period  in the range of $\rm{P} = 12 - 35$ minutes is measured by taking the square root of its power component.
 The amplitude decay of filtered intensities  at a certain  period  as function of distance  along the six paths of  Fig. \ref{AR11416} is fitted with an exponential function of the form
  \begin{eqnarray}
&& A(s)= A_0 exp(-\gamma_d s)+C_0,
\label{fiting}
  \end{eqnarray}
 where, $\gamma_d$ is  the damping length  coefficient,  $  A_0  $ and  $ C_0 $  appropriate constant.
 The projected damping length ($\rm{L_d}=1/\gamma_d$) of filtered  intensity
is calculated  by Matlab Statistics  Toolbox \ Curve  Fitting  and  Distribution  Fitting.
Figure \ref{amplituded}  shows a typical example of the  amplitude decay of  oscillations as a function of distance along loop 1 (blue star signs) with
$\rm{P}=28.43,~27.03,~20.14,~20.34,~13.14 $ and  $13.42~\rm{min}$  that an exponential  function is fitted to the data (red lines) in both passbands (171 and 193 \rm{{\AA}}).
The fitting parameters are  written in the top  of  panels.
Also, in order to compare  the dependence of damping length on its periodicity  in Fig. \ref{amplituded},  the amplitude at each period is normalized to the amplitude of a segment along the path that oscillate with maximum amplitude.
Amplitude decay lengths  of oscillation as a function of period are shown for all six different paths in  Fig. \ref{dampinglengths}. These results indicate that the projected damping lengths
are in the range of 23 Mm to 68 Mm for periods in the rage between 12 and 35 min.
The observed  physical  parameters such as oscillation  period $\rm{P}$, phase speed $\rm{C_s}$, damping length ($\rm{L_d}$), damping time ($\rm{\tau_d=L_{d,est}/C_s}$) and damping quality ($\rm{\tau_d/P}$),
 for filtered oscillation part of intensity  at a sequence of 300 images  with 24 s time intervals are listed in Table\ref{observedresult}  for six different loop segments.
The results of data   analysis   indicate that magnitude of  phase speed  and damping time
along the six different loop segments  are in the range of $38-79~\rm {km~s^{-1}}$ and  $ \rm{\tau_d}\simeq 7- 21~\rm{min}$, respectively.
The   magnitudes of damping quality for dominant oscillation periods in the six  loop segments are in the range of   $ \rm{\tau_d/P}\simeq 0.35- 0.77$   which clearly show that  damping regime is strong for  dominant oscillation periods.

\begin{table*}
\small
\caption{Averages and ranges of physical parameters, such as periodicity, projected damping length, projected  propagation speeds, damping time ($\rm{\tau_d=L_d/C_{s.av}}$), and damping quality (damping time  per period, $\rm{\tau_d/P}$)
of 6 paths along the coronal loops observed with
AIA/SDO 171 and 193 {\AA} passbands.}
\begin{tabular}{@{}|c|r|r|r|r|r|r|r|@{}}
\tableline
\hline
  &\multicolumn{7}{|c|}{ Loop 1}\\
\hline
  $\rm{Band} (\rm{\AA})$ & $\rm{P}(min)$ &\multicolumn{3}{|c|}{$\rm{C_s(km~s^{-1})}$} &$\rm{L_d(Mm)}$ &{$\rm{\tau_d(min)}$}&$\rm{\tau_d/P}$\\ \cline{3-5}

     && $\rm{.min}$ & $\rm{.max}$& $\rm{.av}$& &&\\ [.25ex] 
  \hline
    & 28.43&39.88    & 44.95    &  42.30 & 41.55&    16.37 &  0.58  \\ \cline{2-8}
171 &20.14& 54.97     & 64.51     & 58.52&33.31 & 9.48        & 0.47\\ \cline{2-8}
&13.47&45.18& 47.72& 46.55 & 23.45&    8.40    & 0.62\\ \cline{1-8}
  &27.03&  38.48     &44.99      &41.25& 43.65&   17.63 & 0.65\\ \cline{2-8}
193 &  20.34&47.10& 49.65& 47.51  & 36.72& 12.88  & 0.63 \\ \cline{2-8}
&13.42 &  55.55    &  61.69     &58.40&26.34 &  7.52 & 0.52 \\ \cline{1-8}
\hline
  &\multicolumn{7}{|c|}{ Loop 2}\\
\hline
     &   25.56&39.88     &44.95      &42.30  & 35.08& 13.82 &0.49\\ \cline{2-8}
171 & 20.54 & 48.14     & 53.86     & 50.40 & 32.16& 10.63  &0.53\\ \cline{2-8}
&15.73& 46.47     &50.51      &48.60& 27.14&   9.31  &0.69\\ \cline{1-8}
 & 25.61& 38.48      &39.99     & 39.36 &39.12 &   16.56 &0.61 \\ \cline{2-8}
193 &20.52&  48.24     &31.25      &49.28 &31.32 & 10.60  &0.52 \\ \cline{2-8}
& 15.63&45.92      &50.91     & 48.11&23.54 &  8.15  &0.56\\ \cline{1-8}
\hline
&\multicolumn{7}{|c|}{ Loop 3}\\
\hline
  &  29.44&37.66     &46.33      &42.04 &46.34&18.37&0.62  \\ \cline{2-8}
171&  18.59&47.15      &52.40     & 50.33&35.58&12.77&0.69 \\ \cline{2-8}
&-&-&-&-  &- &-&- \\ \cline{1-8}
  &30.21&37.6&46.30&41.90& 53.05&21.10&0.69 \\ \cline{2-8}
193 &18.74&47.06&53.03&50.01 &43.46&14.48&0.77 \\ \cline{2-8}
&-&-&-&-&-&-&-  \\ \cline{1-8}
\hline
&\multicolumn{7}{|c|}{ Loop 4}\\
\hline
   & 29.20& 51.28      &57.96      &53.95 &33.46&10.34&0.35\\ \cline{2-8}
171  &25.56& 44.67      &48.96      &46.68 &28.03&10.01&0.39\\ \cline{2-8}
 &17.04& 64.36     & 75.02    &  67.57 &23.60&5.82&0.34 \\ \cline{1-8}
& 29.20&39.27&      42.89   &    41.12&35.15&14.25&0.49\\ \cline{2-8}
193&22.71& 44.08       &50.31     & 46.61&30.21&10.80&0.47 \\ \cline{2-8}
& 17.04&  50.12      &55.48      &52.63 &26.56&8.41&0.49 \\ \cline{1-8}
\hline
&\multicolumn{7}{|c|}{ Loop 5}\\
\hline
   &34.07& 51.12     & 51.86      &51.52&44.61 &14.43&0.42 \\ \cline{2-8}
171  &  25.56 &  59.49  &    70.04     & 64.58& 35.28&9.08&0.35  \\ \cline{2-8}
&  20.44& 40.70    & 47.79     & 44.82& 30.19&11.23&0.55\\ \cline{1-8}
     &34.07& 43.43     & 44.72  &   44.29&45.91&17.28& 0.50\\ \cline{2-8}
193&25.56&40.67      &53.71   &  43.92&40.28 &15.28&0.59 \\ \cline{2-8}
 & 20.44&36.78      &40.83     & 38.44 &31.43 &13.62&0.66\\ \cline{1-8}
\hline
&\multicolumn{7}{|c|}{ Loop 6}\\
\hline
   & 29.20&47.11     &22.98    & 59.88 &52.03 &14.48   &
     0.49    \\ \cline{2-8}
171  &25.44& 74.09   &  82.87   &   78.99 &44.10 & 9.30  &
       0.36      \\ \cline{2-8}
& 20.02&  60.30     & 64.68    &  62.68& 31.71&     8.43&
      0.42    \\ \cline{1-8}
      &30.07& 72.88 &     82.94&      77.72 & 67.50&      15.57  &
         0.53    \\ \cline{2-8}
193 &25.56&72.29 &     80.47&      77.29  & 62.45&   13.46   &
        0.52     \\ \cline{2-8}
  & 20.44&58.18      &72.53     & 63.14&27.12 &  7.15&   0.35  \\ \cline{1-8}
\hline
       \end{tabular}
\label{observedresult} 
\end{table*}

\section{ Theoretical considerations }
\begin{figure}    
 \centerline{\includegraphics[width=.5\textwidth,clip=]{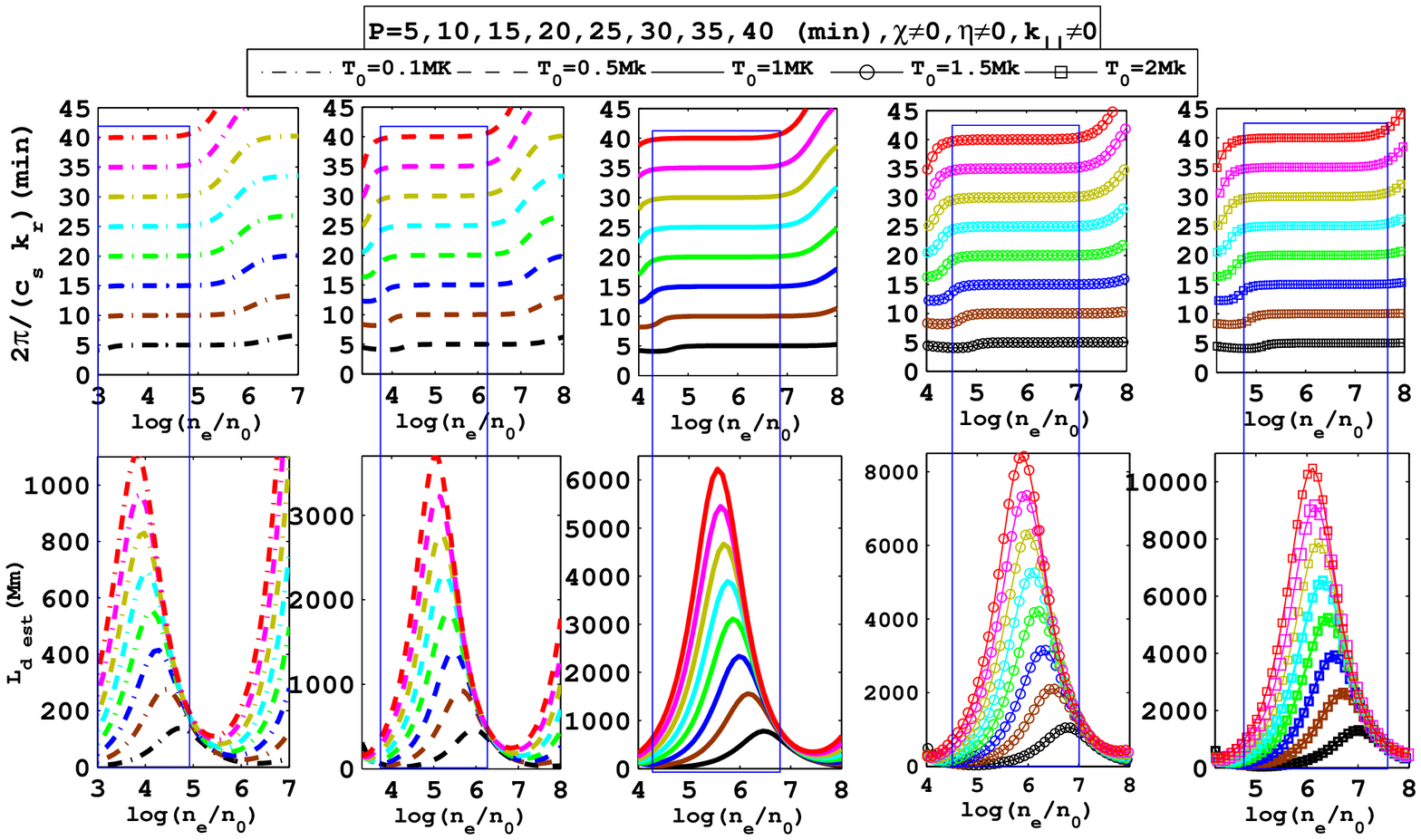}
              }
              \caption{Plot of the the estimated damping length $\rm{L_{d,est}}$  in Mm (bottom), and $2\pi/(k_r \rm{C_s})$ in min as a  function of the dimensionless electron number density $(\rm{log(n_e/n_0),~ n_0=10^{4}cm^{-3}})$ (top)  due to the presence  of all three dissipation mechanisms
           for different temperatures $\rm{T_0=0.1, 0.5, 1, 1.5~and~2}$ MK and  periods $\rm{P=5, 10, 15, 20, 25, 30, 35~and~40} $ min in the acceptable range. The acceptable range of damping length and number density  are outlined with blue rectangle   in the panels.}
             \vspace{-0.02\textwidth}
   \label{chietaka3}
      \end{figure}

Theoretical studies investigating the damping of the slow wave have concentrated on the effects
of thermal conduction, compressive viscosity, optically thin radiation, gravitational stratification and magnetic field divergence.
Generally, thermal conduction is found to be the dominant mechanism for dissipation of  slow modes in the solar corona
(De Moortel and Hood, \citeyear{ DeMoortelh2003}, \citeyear{DeMoortelh2004a}, \citeyear{DeMoortel2009}; Pandey and  Dwivedi  \citeyear{pandey2006}; Ofman and Wang, \citeyear{OfmanW2002}; Van Doorsselaere et al. \citeyear{VanDoorsselaere2011}; Abedini et al. \citeyear{Abedini2012}).
Here, the coronal loops  are  considered  a homogeneous plasma medium
in the presence of thermal conduction,  compressive viscosity and optically thin radiation  dissipation mechanism
 with constant equilibrium values $\rho_0 $, $ \rm{T_0} $, $ p_0 $  and no flow ($ v_0=0 $), also with  a uniform background magnetic field along the loops.
Assuming all disturbances in terms of Fourier components for frequency ($\omega$)  and wave number ($k$)
 in z-direction, $ \exp i(k z-\omega t) $, combining linearized MHD equations in the presence of
 thermal conduction $(E_c=\frac{\partial}{\partial z}(k_{||}\frac{\partial \rm{T_0}}{\partial z}) )$, compressive viscosity $(E_\eta=\frac{4}{3}\eta_0(\frac{\partial v}{\partial z})^2 )$ and
   optically thin radiation  $(E_r=\chi \rm{n_e^2} \rm{T_0}^{\alpha})$ (Sigalotti et al. \citeyear{Sigalotti2007}) leads to the following dispersion relation:
 \begin{eqnarray}\label{dispersion.rr}
&& i{\cal A}k^4+i{\cal B}k^2+i{\cal C}=0,\\
&& {\cal A}=-(\gamma - 1)k_{||}[\frac{4}{3}\frac{\eta_0 \rm{T_0}}{\rho_0 p_0}\omega+ \frac{\rm{T_0}}{\rho_0}],\nonumber\\
&& {\cal B}=(\gamma -1)[k_{||}\frac{\rm{T_0}}{p_0}\omega^2+\frac{4}{3}\frac{\eta_0}{\rho_0}+
i\omega\frac{\gamma p_0}{\rho_0},\nonumber\\
&&~~~~~+\chi(i\omega\frac{4}{3}\alpha  \frac{ \eta_0}{p_0}-\alpha+2)\frac{\rho_0  \rm{T_0}^\alpha}{\overline{m}^{2}}],\nonumber\\
&& {\cal C}=\omega^2[\alpha\chi(\gamma - 1)\frac{\rho_0^2 \rm{T_0^\alpha}}{\overline{m}^{2}p_0}-i\omega],
\nonumber
  \end{eqnarray}
Here, $\gamma=5/3$, $\overline{m}=0.6 m_{H}$, $ \eta_{0} = 10^{-17}T_{0}^{ 5/2}  \rm{k g m^{-1}s^{-1}}$,
$ k_{||} = 10^{-11}\rm{T_{0}^{ 5/2}}~\rm{W m^{-1}K^{-1}}$ are taken and $\chi$ and  $\alpha $ are temperature dependent ( Braginskii, \citeyear{Braginskii1965}; Hildner,~\citeyear{Hildner1974}).
Furthermore, the $ \omega $   is assumed to be real and the wave number is imaginary ($k=k_r+ik_i $).
The measured propagation speed of the intensity disturbances is
 a projected component of sound speed which is perpendicular to the LOS (line-of-sight).
 By analyzing the filtered intensity with particular periods in the range of ($ 8-40~\rm{min}) $,
 the average value of projected sound speed  along the selected paths was found in the range of $ 38-79 ~\rm{km~s^{-1}} $.
 So, it can be concluded that average value of real sound speed  is greater than   $ 38~\rm{km~s^{-1}} $.
 The temperature associated with  $ \rm{C_s}=38~\rm{km~s^{-1}} $ is found  $ 0.1~\rm{MK} $
by using the theoretical relation  $ T_{0}=\frac{\overline{m}C_s^2}{\gamma k_{B}}$,
and by inserting the  Boltzmann constant $k_B = 1.38\times10^{-23}~\rm{JK^{-1}}$.
In a propagating  wave with a specific oscillation period $\rm{P}$ that has a damping length $\rm{L_d}$, it is expected to have $\rm{P}=2\pi/(k_r \rm{C_s})$, and $\rm{L_d}= 1/k_i $.
 In order to estimate damping length, the wave number of propagation wave associated with  oscillation period
 ($ 5- 40\rm{min} $) at different values of temperature ($ \rm{T_0=0.1,~0.5,~1,~1.5,~and` ~2~\rm{MK} }$)
   is calculated   based on numerical solutions  of  dispersion relation (\ref{dispersion.rr}).
 Plot of   $2\pi/(k_r \rm{C_s})$ in (min) (top panels) and the estimated  damping length $\rm{L_{d,est}}=1/k_i $  in Mm (bottom panels), as a  function of $(\rm{log(n_e/n_0),~ n_0=10^{4}~cm^{-3}})$  due to the presence
of  combining dissipation mechanism are shown in Fig. \ref{chietaka3}
for different values of temperature  and  oscillation period.
In  each panel of  Fig. \ref{chietaka3}, acceptable range ($2\pi/(k_r \rm{C_s)=5,~10,~15,~20,~25,~30,~35~and~40}$ min) of damping length and number density  are outlined with blue rectangle.
An interesting prediction of this model is that the plots of $\rm{L_{d,est}} $ as a  function of $(\rm{log(n_e/n_0)})$ have a maximum in  acceptable range.
Another interesting feature visible in these plots is
that both   damping lengths and their  maximum value  increase  with increasing $\rm{T_0}$ and $\rm{P}$. Also, the maximum point of $\rm{L_{d,est}} $  shifts toward bigger electron number density.
 These results  reveal that treatment  of  estimated damping due to the presence of dissipation mechanism  is complex and its value depends not only on the oscillation period but also on the electron number density and temperature.
 Moreover, the results  based on numerical solutions of dispersion relation illustrate that,
 although the damping length and damping time can be increased
with increasing the oscillation periods, the dependence on periodicity  and value of
these quantities strongly depend on the temperature and the electron number
density.

\begin{table*}
\small
\caption{Ranges  of physical parameters such as estimated damping length ($\rm{L_{d,est}=1/k_i}$), damping time ($\rm{\tau_d=L_{d,est}/C_s}$ ), damping quality ($\rm{\tau_{d,est}/P}$) and electron number density within
the acceptable   regions ( $2\pi/(k_r \rm{C_s)=5,10,...,35,~and~40}$  min) are calculated at  dominant oscillation  periods
     for different values of  temperature ($ \rm{T_0=0.1, 0.5, 1, 1.5~and~2~\rm{MK} }$).}
      \begin{tabular}{@{}|c|r|r|r|r|r|r|r|@{}}
\tableline
\hline
  $\rm{T_0(MK)}$& $\alpha$ & $\chi$ &\rm{P(min)}&\multicolumn{4}{|c|}{ $\chi\neq0,\eta\neq0,k_{||}\neq0$}\\ \cline{4-8}
    &&&& $\rm{ log(n_e/n_0)}$ & $\rm{L_{d,est}(Mm)}$& $\rm{\tau_{d,est}(min)}$ & $\rm{\tau_{d,est}/P}$\\ [.25ex] 
       &&&& Range & Range & Range & Range\\ [.25ex] 
   \hline
   &&& \rm{5 }  & 3-5.8  &  $\geq 9$   &$ \geq4 $ &  $\geq 0.66  $ \\ \cline{4-8}
  &&& \rm{10 } & 3-5.6   & $\geq 19$   & $\geq 8.5 $&  $\geq  0.72  $   \\ \cline{4-8}
 && &  \rm{15} &3-5.4  & $\geq 48$ & $ \geq20.5  $ &  $\geq   1.16   $\\  \cline{4-8}  
0.1&0& $8\times10^{-34}$ & \rm{20 }  & 3-5.1  &  $\geq 80$   &$\geq35 $ &  $\geq  1.49    $ \\ \cline{4-8}
   &&& \rm{25 } & 3-5   & $\geq 134$   &$ \geq54.5$   &  $\geq 1.85 $   \\ \cline{4-8}
  & & & \rm{30 } &3-4.95  & $\geq 180$ &$\geq78  $ &  $\geq   2.22  $\\  \cline{4-8}
 & &&\rm{35 }  & 3-4.85  &  $\geq 230$   &$\geq97 $ &  $\geq  2.36 $ \\ \cline{4-8}
   &&& \rm{40} & 3-4.75   & $\geq 332$   &$\geq138$&  $\geq  2.93$   \\ \cline{1-6}
  \hline
   &&& \rm{5 }  & 3.4-7.8 &  $\geq 18$   & $\geq   3   $&  $\geq 0.56 $ \\ \cline{4-8}
  & &&\rm{10 } &3.4-7.6  & $\geq 36$   &$\geq  6 $&  $\geq    0.58 $   \\ \cline{4-8}
 &   &&\rm{15} & 3.4-7.4 & $\geq 58$ &$\geq   8.9   $&  $\geq    0.59 $\\  \cline{4-8}  
0.5& -2.5&$3.94\times10^{-21}$& \rm{20 }  &  3.4-7  &  $\geq 76$   &$\geq 12  $&  $\geq    0.6  $ \\ \cline{4-8}
   && &\rm{25 } & 3.4-6.8  & $\geq 150$   &$\geq  24.5  $&  $\geq   0.93 $   \\ \cline{4-8}
  &&   &\rm{30 } &3.4-6.6  & $\geq 221$ &$\geq  36   $&  $\geq    1.14 $\\  \cline{4-8}
 &&& \rm{35 }  & 3.4-6.4   &  $\geq 332$   &$\geq 54  $&  $\geq    1.47   $ \\ \cline{4-8}
   &&& \rm{40} & 3.4-6.25    & $\geq 385$   &$\geq  63$&  $\geq  1.49$   \\ \cline{1-6}
     \hline
   & &&\rm{5 }  &  4.7-8.2  &  $\geq 25$   &$\geq   2.75  $&  $\geq   0.53 $ \\ \cline{4-8}
  && &\rm{10 } & 4.6-8 & $\geq 50$   &$\geq   5.54  $&  $\geq    0.53 $   \\ \cline{4-8}
 & &  &\rm{15} & 4.5-7.8 & $\geq 80$ &$\geq   8.86 $&  $\geq    0.57  $\\  \cline{4-8}  
1 &-1&$5.51\times10^{-30}$& \rm{20 }  &  4.4-7.5 &  $\geq 130$   &$\geq   14.4 $&  $\geq  0.69 $ \\ \cline{4-8}
   &&& \rm{25 } & 4.3-7.3    & $\geq 210$   &$\geq   23.25 $&  $\geq    0.89 $   \\ \cline{4-8}
  & &&  \rm{30 } & 4.25-7.1   & $\geq 320$ &$\geq   35.43 $&  $\geq    1.13  $\\  \cline{4-8}
 &&& \rm{35 }  &4.2-6.95    &  $\geq 380$   &$\geq  42.08 $&  $\geq   1.15 $ \\ \cline{4-8}
   &&& \rm{40} & 4.25-6.75  & $\geq 540$   &$\geq   59.79$&  $\geq   1.43$   \\ \cline{1-6}
   \hline
   &&& \rm{5 }  & 5-8  &  $\geq 30$   &$\geq  2.7  $&  $\geq     0.52 $ \\ \cline{4-8}
  &&& \rm{10 } & 4.8-8   & $\geq 62$   &$\geq   5.6 $&  $\geq    0.54  $   \\ \cline{4-8}
 &&  & \rm{15} & 4.6-8 & $\geq 126$ &$\geq  11.4 $&  $\geq   0.73  $\\  \cline{4-8}  
1.5& -1&$5.51\times10^{-30}$& \rm{20 }  & 4.5-7.8 &  $\geq 158$   &$\geq   14.3 $&  $\geq   0.69 $ \\ \cline{4-8}
   &&& \rm{25 } &4.7-7.6    & $\geq 240$   &$\geq   21.7  $&  $\geq    0.83 $   \\ \cline{4-8}
  & & & \rm{30 } &4.6-7.4  & $\geq 360$ &$\geq 32.6 $&  $\geq    1  $\\  \cline{4-8}
 &&& \rm{35 }  &4.5-7.2   &  $\geq 406$   &$\geq   36.7 $&  $\geq   1 $ \\ \cline{4-8}
   &&& \rm{40} & 4.4-7     & $\geq 590$   &$\geq   53.4$&  $\geq    1.3$   \\ \cline{1-6}
   &&& \rm{5 }  & 5.2-8 &  $\geq 37$   &$\geq 2.89 $&  $\geq  0.55 $ \\ \cline{4-8}
  &&& \rm{10 } &  5-8  & $\geq 71$   &$\geq    5.54 $&  $\geq    0.53 $   \\ \cline{4-8}
 & &&  \rm{15} & 4.8-8 & $\geq 112$ &$\geq    8.74 $&  $\geq    0.56 $\\  \cline{4-8}  
2 &-1 &$5.51\times10^{-30}$&\rm{20 }  & 4.4-8  &  $\geq 180$   &$\geq  14$&  $\geq    0.67  $ \\ \cline{4-8}
   & &&\rm{25 } & 4.8-8   & $\geq 350$   &$\geq    27.3   $&  $\geq   1 $   \\ \cline{4-8}
  &   &&\rm{30 } & 4.8-8 & $\geq 440$ &$\geq 34.32  $&  $\geq    1 $\\  \cline{4-8}
 & &&\rm{35 }  &  4.8-7.8 &  $\geq 532$   &$\geq  41.5  $&  $\geq    1.14$ \\ \cline{4-8}
   &&& \rm{40} &  4.75-7.6  & $\geq 610$   &$\geq  47.58$&  $\geq 1.14$   \\ \cline{1-6}
    \hline
   \end{tabular}
\label{tresults} 
\end{table*}
\section{ Discussion and Conclusions}
The aim of this paper is to  study the physical parameters  of longitudinal intensity
variations in the large active-region loop systems by AIA/SDO.
So, 300 high  cadence  images  of loop systems  with an interval of 24 seconds in the 171$\rm{\AA}$ and 193 $\rm{\AA}$  are analyzed.
First, the intensity  as a function of distance along the six loop segments  is calculated for all times.
Then a background constructed from the 100 point (40 min) running average in time is
subtracted from the original intensities.
For example, the time-distance maps of intensity along the loop 1 before (panels a) and b )) and after (panels  c) and d)) removal  of the background intensity are shown in top  and bottom rows of  Fig. \ref{Iandi}, respectively.
The Fourier  power spectra and period-distance map of fluctuating part of intensities along the loop segments
reveal periods in the range of $\rm{P}\simeq 8-40$ minutes for both the AIA channels (for example, see the  middle and bottom rows  of Fig. \ref{Iandi}).
Moreover, the filtered fluctuating intensity of loops  shows   small amplitude periodic variations which are interpreted as evidence for   damping of slow  magneto-acoustic waves in loop systems
(see bottom panels in Fig. \ref{Iandi} and  Fig. \ref{phasespeed}).
This range of oscillation period that has been estimated in this study comfortably overlaps with values quoted by other authors who analyzed similar observations of propagating fluctuations and  MHD theory's predictions for the  slow wave (see, {\it e.g.}, De Moortel et al. \citeyear{DeMoortel2002}, \citeyear{DeMoortelh2003}, \citeyear{DeMoortel2004}; Marsh et al. \citeyear{Marsh2003}, \citeyear{marsh2004}, \citeyear{Marshw2009}, \citeyear{Marsh2011}; Krishna  et al. \citeyear{Krishna2014}, Yuan and Nakariakov \citeyear{Yuan2012}, Kiddie et al. \citeyear{Kiddie2012}).
Time-distance images formed from the integrated intensity along the loop segments  show  that
disturbances propagate approximately with a constant speed.
Accordingly, the projected speeds of  filtered intensity  along  the 6 loop segments   are estimated   for  some dominant oscillation period between 12 min and 35 min, by fitting  a linear function  to the points with maximum amplitude fluctuation of time-distance maps.
 Also, the  damping length ($\rm{L_d}$) of oscillation are calculated by fitting  an exponentially  decreasing function to the filtered fluctuation part of intensity profiles (see Fig. \ref{amplituded}).
Other physical parameters, such as damping time ($\rm{\tau_d=L_d/C_{s.av}}$) and damping quality ($\rm{\tau_d/P}$), are calculated  by using the speeds and damping lengths.
The some results of analysis  in the 171 {\AA} and 193 {\AA} passbands   are listed in Table \ref{observedresult}
for different loop segments.
 These observed  results show that:
  \begin{itemize}
\item The average values of  projected speeds
 for  dominant oscillation periods of six different loop segments are  in the range of  $38-79~ \rm{km~s^{-1}}$.
 \item The average values of  damping lengths and damping times  are  in the range of  $23 -68~ \rm{Mm}$ and
    $  7- 21~\rm{min}$, respectively.
  Also, its values are  sensitive to the oscillation period and  increase  with increasing $\rm{T_0}$ and $\rm{P}$.
  \item The magnitude value of damping qualities   are obtained  in the range of $ <1$  for filtered intensity  which  correspond to strong damping regime.
 \end{itemize}
Also, to compare observed results  with the  theoretical MHD prediction,
the estimated physical parameters  of the slow wave that  has concentrated on  the effects of thermal conduction, compressive viscosity and optically thin radiation is investigated.
The damping lengths of filtered intensities  as function of electron number density based on the theoretical MHD prediction  are shown in Fig. \ref{chietaka3}  for different values of temperature ($ \rm{T_0=0.1,~ 0.5,~ 1.5~and~2~\rm{MK}} $).
The magnitudes of estimated damping lengths and other physical parameters  in the  acceptable range ($2\pi/(\rm{C_s} k_r) \simeq \rm{5,10,...,35,~and~40~\rm{min}}$) are listed in Table  \ref{tresults}.
 These  results show that:
\begin{itemize}
    \item In  the presence of three dissipation mechanisms  together
             damping lengths and damping times increase with increasing temperature  and oscillation period but  damping qualities decrease with increasing temperature (see Table \ref{tresults})
    \item  The plots of damping length as  function of electron number density have a maximum in  acceptable range (see  Fig. \ref{chietaka3}).
    \item The damping lengths and their maximum value  increase  with increasing $\rm{T_0}$ and $\rm{P}$. Also, the maximum point of plots  shift toward bigger electron number density.
    \item The dependence on periodicity  and value of damping length strongly depend  on the temperature and the electron number density.
  These theoretical results  reveal that treatment  of  estimated damping due to the presence of all three dissipation mechanisms  is complex and its value depends not only on the oscillation  period but also  on the electron number density and temperature.
 \end{itemize}
 The magnitudes of observed damping times and damping qualities  are independent of the LOS.
 comparing the observed damping times and dependence of  damping times on their periods with the theoretically predicted values, it can be seen  at  low  periods ($\rm{P}<15~ \rm{min}$), the damping times   and  the dependence of  damping times  on their periods of  propagating slow magneto-acoustic waves in the gravitationally stratified loops that are situated
above active regions   may be explained by considering  three dissipation mechanisms in an especial
 range of electron number density ($\rm{n_e}\simeq 10^{7}-10^{12} ~\rm{cm^{-3}}$) but at high periods  the other dissipation mechanism   must be  considered.
 Also, It can be concluded that  the behavior of  propagating slow magneto-acoustic waves along the corona loops,
 in addition to the dispersion mechanism, can depend on the other quantities such as temperature, electron number density, and so on.
\begin{acknowledgements}
I thank the anonymous referee,  Dr. Mike Marsh and Dr. H. Safari  for the very helpful comments and suggestions.
\end{acknowledgements}


\begin{thebibliography}{}
\bibitem[\protect\citeauthoryear{Abedini}{2011}]{Abedini2011}
Abedini, A., Safari, H.: \na,~\textbf{16}, 317 (2011)
\bibitem[\protect\citeauthoryear{Abedini}{2012}]{Abedini2012}
 Abedini, A., Safari, H., Nasiri, S.: \solphys,~\textbf{280}, 137 (2012)
 \bibitem[\protect\citeauthoryear{Banerjee}{2007}]{Banerjee2007}
 Banerjee, D., Erd\'{e}lyi, R., Oliver, R., \'{O}shea, E.: \solphys,~\textbf{246}, 3 (2007)
 \bibitem[\protect\citeauthoryear{Braginskii}{1965}]{Braginskii1965}
  Braginskii, S.I.: Rev, Plasma Phys.,~\textbf{1}, 205 (1965)
 \bibitem[\protect\citeauthoryear{Berghmans}{1999}]{Berghmans1999}
 Berghmans, D., Clette, F.: \solphys,~\textbf{186}, 207 (1999)
 \bibitem[\protect\citeauthoryear{Deforest}{1998}]{Deforest1998}
DeForest, C.E., Gurman, J.B.: \apj,~\textbf{501}, L217 (1998)
 \bibitem[\protect\citeauthoryear{DeMoortel}{2000}]{DeMoortel2000}
De Moortel, I., Ireland, J., Walsh, R.W.: \aap,~\textbf{355}, L23 (2000)
 \bibitem[\protect\citeauthoryear{DeMoortel}{2002}]{DeMoortel2002}
De Moortel, I., Ireland, J., Walsh, R.W., Hood, A.W.: \solphys,~\textbf{209}, 61 (2002)
\bibitem[\protect\citeauthoryear{DeMoortel}{2003}]{DeMoortelh2003}
De Moortel, I., Hood, A.W.: \aap,~\textbf{408}, 755 (2003)
\bibitem[\protect\citeauthoryear{DeMoortel}{2004a}]{DeMoortelh2004a}
De Moortel, I., Hood, A.W.: \aap,~\textbf{415}, 705 (2004a)
 \bibitem[\protect\citeauthoryear{DeMoortel}{2004}]{DeMoortel2004}
De Moortel, I., Hood, A.W., De Pontieu, B.: ESASP,~\textbf{547}, 427 (2004)
\bibitem[\protect\citeauthoryear{DeMoortel}{2009}]{DeMoortel2009}
De Moortel, I.: \ssr,~\textbf{149}, 65 (2009)
\bibitem[\protect\citeauthoryear{Erdelyi}{2008}]{Erdelyi2008}
Erd\'{e}lyi, R., Luna-Cardozo, M., Mendoza-Brice\~{n}o, C.A.: \solphys,~\textbf{252}, 305 (2008)
\bibitem[\protect\citeauthoryear{Hildner}{1974}]{Hildner1974}
 Hildner, E.: \solphys,~\textbf{35}, 123 (1974)
  \bibitem[\protect\citeauthoryear{Kiddie}{2012}]{Kiddie2012}
Kiddie, G., De Moortel, I., Del Zanna, G., McIntosh, S.W., Whittaker, I.: \solphys,~\textbf{279}, 427 (2012)
\bibitem[\protect\citeauthoryear{Klime}{2002}]{Klieme2002}
 Kliem, B., Dammasch, I.E., Curdt, W., Wilhelm, K.: \apj,~\textbf{568}, L61 (2002)
\bibitem[\protect\citeauthoryear{Klimchuk}{2004}]{Klimchuk2004}
Klimchuk, J.A., Tanner, S.E.M., De Moortel, I.: \apj,~\textbf{616}, 1232 (2004)
\bibitem[\protect\citeauthoryear{Krishna}{2014}]{Krishna2014}
Krishna Prasad, S., Banerjee, D., Van Doorsselaere, T.: \apj,~\textbf{789}, 118 (2014)
\bibitem[\protect\citeauthoryear{Mariska}{2006}]{Mariska2006}
Mariska, J.T.: \apj,~\textbf{639}, 484 (2006)
\bibitem[\protect\citeauthoryear{Mariska}{2010}]{Mariska2010}
 Mariska, J.T., Muglach, K.: \apj,~\textbf{713}, 573 (2010)
 \bibitem[\protect\citeauthoryear{Marsh}{2003}]{Marsh2003}
 Marsh, M.S., Walsh, R.W., De Moortel, I., Ireland, J.: \aap,~\textbf{404}, L37 (2003)
 \bibitem[\protect\citeauthoryear{Marsh}{2004}]{marsh2004}
Marsh, M.S., Walsh, R.W., De Moortel, I., Ireland, J.: ESASP.,~\textbf{547}, 519 (2004)
\bibitem[\protect\citeauthoryear{Marshw}{2009}]{Marshw2009}
Marsh, M.S.,  Walsh, R.W., Plunkett, S.: \apj,~\textbf{697}, L1674 (2009)
\bibitem[\protect\citeauthoryear{Marsh2011}{2011}]{Marsh2011}
 Marsh, M.S., De Moortel, I., Walsh, R.W.: \apj,~\textbf{734}, 81 (2011)
 \bibitem[\protect\citeauthoryear{McEwan}{2006}]{McEwan2006}
McEwan, M.P., De Moortel, I.: \aap,~\textbf{448}, 763 (2006)
\bibitem[\protect\citeauthoryear{Mendoza}{2004}]{Mendoza2004}
Mendoza-Brice\~{n}o, C.A., Erd\'{e}lyi, R., Sigalotti, L., Di, G.: \apj,~\textbf{605}, 493 (2004)
\bibitem[\protect\citeauthoryear{Nakariakov}{2000}]{Nakariakov2000}
Nakariakov, V.M., Verwichte, E., Berghmans, D., Robbrecht, E.: \aap,~\textbf{362}, 1151 (2000)
\bibitem[\protect\citeauthoryear{Nakariakov}{2005}]{Nakariakov2005}
Nakariakov, V.M., Verwichte, E.: LRSP, \textbf{2}, 3 (2005)
 \bibitem[\protect\citeauthoryear{Ofman}{1997}]{Ofman1997}
Ofman, L., Romoli, M., Poletto, G., Noci, G., Kohl, J.L.: \apj,~\textbf{491}, L111 (1997)
 \bibitem[\protect\citeauthoryear{Ofman}{1999}]{Ofman1999}
Ofman, L., Nakariakov, V.M., DeForest, C.E.: \apj,~\textbf{514}, 441 (1999)
 \bibitem[\protect\citeauthoryear{Ofman}{2000}]{Ofman2000}
 Ofman, L., Nakariakov, V.M., Sehgal, N.: \apj,~ \textbf{ 533}, 1071 (2000)
 \bibitem[\protect\citeauthoryear{Ofman}{2002}]{OfmanW2002}
 Ofman, L., Wang, T.J.: \apj, ~\textbf{580}, L85 (2002)
\bibitem[\protect\citeauthoryear{pendy}{2006}]{pandey2006}
 Pandey, V.S., Dwivedi, B.N.: \solphys,~\textbf{236}, 127 (2006)
\bibitem[\protect\citeauthoryear{porter}{1994}]{porter1994a}
Porter, L.J., Klimchuk, J.A., Sturrock, P.A.: \apj,~\textbf{435}, 502 (1994)
\bibitem[\protect\citeauthoryear{Roberts}{2000}]{Roberts2000}
 Roberts, B.: \solphys,~\textbf{193}, 139 (2000)
\bibitem[\protect\citeauthoryear{Roberts}{2006}]{Roberts2006}
Roberts, B.: RSPTA.,~\textbf{364}, 447 (2006)
\bibitem[\protect\citeauthoryear{Sakurai}{2002}]{Sakurai2002}
 Sakurai, T., Ichimoto, K., Raju, K.P., Singh, J.: \solphys,~\textbf{209}, 265 (2002)
 \bibitem[\protect\citeauthoryear{Sigalotti}{2007}]{Sigalotti2007}
  Sigalotti, L., Di, G., Mendoza-Brice\~{n}o, C.A.,
 Luna-Cardozo, M.: \solphys,~\textbf{246}, 187 (2007)
 \bibitem[\protect\citeauthoryear{Sych}{2014}]{Sych2014}
 Sych, R., Nakariakov, V. M.: \aap,~\textbf{569}, A72 (2014)
 \bibitem[\protect\citeauthoryear{Threlfall2013}{2013}]{Threlfall2013}
  Threlfall, J., De Moortel, I., McIntosh, S.W., Bethge, C.: \aap,~\textbf{556}, A124 (2013)
 \bibitem[\protect\citeauthoryear{Tsiklauri}{2001}]{Tsiklauri2001}
Tsiklauri, D., Nakariakov, V.M.: \aap,~\textbf{379}, 1106 (2001)
\bibitem[\protect\citeauthoryear{Uchida}{1968}]{Uchida1968}
  Uchida, Y.: \solphys,~\textbf{4}, 30 (1968)
  \bibitem[\protect\citeauthoryear{Uchida}{1970}]{Uchida1970}
  Uchida, Y.: PASJ.,~\textbf{22}, 341 (1970)
\bibitem[\protect\citeauthoryear{Uritsky}{2013}]{Uritsky2013}
Uritsky, V.M., Davila, J.M., Viall, N.M., Ofman, L.: \apj,~\textbf{778}, 26 (2013)
\bibitem[\protect\citeauthoryear{VanDoorsselaere}{2011}]{VanDoorsselaere2011}
 Van Doorsselaere, T., Wardle, N., Del Zanna, G., Jansari, K., Verwichte, E., Nakariakov, V.M.:
  \apj,~\textbf{727}, L32 (2011)
  \bibitem[\protect\citeauthoryear{Verwichte}{2005}]{VerwichteN2005}
 Verwichte, E., Nakariakov, V.M., Cooper, F.C.: \aap,~\textbf{430}, L65 (2005)
\bibitem[\protect\citeauthoryear{Verwichte}{2008}]{Verwichte2008}
 Verwichte, E., Haynes, M., Arber, T.D., Brady, C.S.: \apj,~\textbf{685}, 1286 (2008)
 \bibitem[\protect\citeauthoryear{Wang}{2002a}]{Wang2002a}
Wang, T.J., Solanki, S.K., Curdt, W., Innes, D.E., Dammasch, I.E.: \apj,~\textbf{574}, L101 (2002a)
\bibitem[\protect\citeauthoryear{Wang}{2002c}]{Wang2002c}
Wang, T.J., Solanki, S.K., Curdt, W., Innes, D.E., Dammasch, I.E.: ESASP.,~\textbf{505}, 199 (2002c)
\bibitem[\protect\citeauthoryear{Wang}{2003a}]{Wang2003a}
Wang, T.J., Solanki, S.K., Innes, D.E., Curdt, W., Marsch, E.: \aap,~\textbf{402}, L17 (2003a)
\bibitem[\protect\citeauthoryear{Wang}{2003b}]{Wang2003b}
Wang, T.J., Solanki, S.K., Curdt, W., Innes, D.E., Dammasch, I.E., Kliem, B.: \aap,~\textbf{406}, 1105 (Paper I) (2003b)
\bibitem[\protect\citeauthoryear{Wang}{2009}]{Wang2009}
 Wang, T.J., Ofman, L., Davila, J.M.,  Mariska, J.T.: \aap,~\textbf{503}, L25 (2009)
\bibitem[\protect\citeauthoryear{Wang}{2011}]{Wang2011}
 Wang, T.J.: \ssr,~\textbf{158}, 397 (2011)
\bibitem[\protect\citeauthoryear{Yuan}{2012}]{Yuan2012}
Yuan, D., Nakariakov, V.M.: \aap,~\textbf{543}, A9 (2012)
\end{thebibliography}
\end{document}